\documentclass[final,OA]{AAAI-Std}

\def\artid{0001}%

\usepackage{tabularx}
\usepackage{subfiles}

\makeatletter
\def\ps@preprint{%
  \def\@oddhead{\normalfont \small PREPRINT\hfil\thepage}%
  \let\@evenhead\@oddhead
  \let\@oddfoot\@empty
  \let\@evenfoot\@empty
}
\makeatother

\begin{document}

\title{Goals as a Representational Layer for Human--AI Collaboration}

\author[1]{Gaia Molinaro}
\author[2]{Lev Tankelevitch}
\author[2]{Sean Rintel}
\author[1]{Scott Counts}

\address{\textsuperscript{1}Microsoft; \textsuperscript{2}Microsoft Research.}

\abstract[Abstract]{Goals are central to how people, teams, and organizations structure their work. Yet, while generative artificial intelligence (AI) systems increasingly operate across a worker's context, information about \textit{why} a person is pursuing a task, beyond local prompts and interactions, is typically missing. We argue that goals should become an explicit representational layer mediating interactions among people, AI systems, and organizations. Because human goals are hierarchical, partially latent, and dynamically evolving, such a layer must be multidimensional, both grounded in human articulation and inferred from available evidence, and continuously updated. A goal layer could connect immediate requests to broader outcomes, improve the organization and retrieval of relevant context, support reflection and coordination, and provide a foundation for training, aligning, and evaluating AI systems in terms of human-centered outcomes. We highlight these opportunities while identifying open challenges involving goal inference, socio-cognitive frictions, alignment, privacy, and governance.}

\maketitle
\pagestyle{preprint}
\thispagestyle{preprint}

\section{Introduction} 
Generative artificial intelligence (AI) is rapidly changing how people work, becoming integrated into existing activities while enabling new ones \citep{counts2026ai, dillon2025shifting, handa2025economic}. To support this expanding range of work, AI systems increasingly draw on contextual information from users' artifacts and interactions, as well as from organizational resources \citep{lewis2020retrieval, ram2023context}. However, information about \emph{why} a worker is pursuing a task\,---\,the \emph{goals} that motivate, energize, and direct it\,---\,is typically missing. Even when users fully specify an immediate request, the larger objectives it serves often remain latent. In most cases, goals remain implicit, scattered across multiple surfaces, or communicated verbally without any persistent representation. 

 The difficulty of representing goals predates the AI era, as longstanding approaches such as SMART goals and OKRs attest \citep{doran1981smart,grove2015high}. AI-mediated work, however, makes the problem more consequential. Human collaborators can often recover unspoken objectives through shared experience, tacit knowledge, and opportunities for clarification; AI systems have less reliable access to these cues \citep{mathur2024advancing}. Indeed, several forms of misalignment characterize issues in human--AI collaboration. Users experience friction as they repeatedly reconstruct their intent \citep{chen2026conversation}. Agents drift away from the original objective in multi-turn interactions \citep{arike2025evaluating, laban2026llms, park2026authorship} and struggle to support users with evolving goals \citep{tack2026llms}. The few approaches attempting to remedy these issues tend to focus on localized user intents, rather than broader objectives that motivate current efforts \citep{coscia2025ongoal, kim2026discoverllm, kim2026intentflow, wu2025collabllm}. At the same time, people are delegating increasingly complex and consequential work to AI \citep{anthropic2026aeiv6, sarkar2023exploring}. As a result, an existing human coordination problem turns into a human--AI coordination challenge, with greater scope for actions to diverge from the outcomes users intend and more severe safety and security threats \citep{jia2025consequences, mitchell2025fully, passi2025agentic}. 

Goals\,---\,mental representations of desired future states \citep{elliot2008goal}\,---\,are particularly well suited to addressing this gap because they already provide an organizing principle for human behavior and collaborative work. Goals direct attention and effort, provide standards for evaluating progress, guide decisions among possible courses of action, and enable flexible coordination  \citep{molinaro2023goal, keeney1976decision}. This organizing role is particularly relevant to knowledge work, where goals direct both actions and decisions \citep{keeney1976decision, reinhardt2011knowledge, tankelevitch2026goals} while igniting motivation and driving performance \citep{locke2002building}. In collaborative settings, goals provide a basis for coordinating interdependent activities and establishing shared direction \citep{chen2025track, devisch2020shared, tankelevitch2026nudging, scott2024mental}. Goals also matter for workers' well-being: making progress on meaningful work is associated with motivation and positive affect \citep{amabile2011progress}. 

\begin{figure*}[ht!]
    \centering
    \includegraphics[width=1\linewidth]{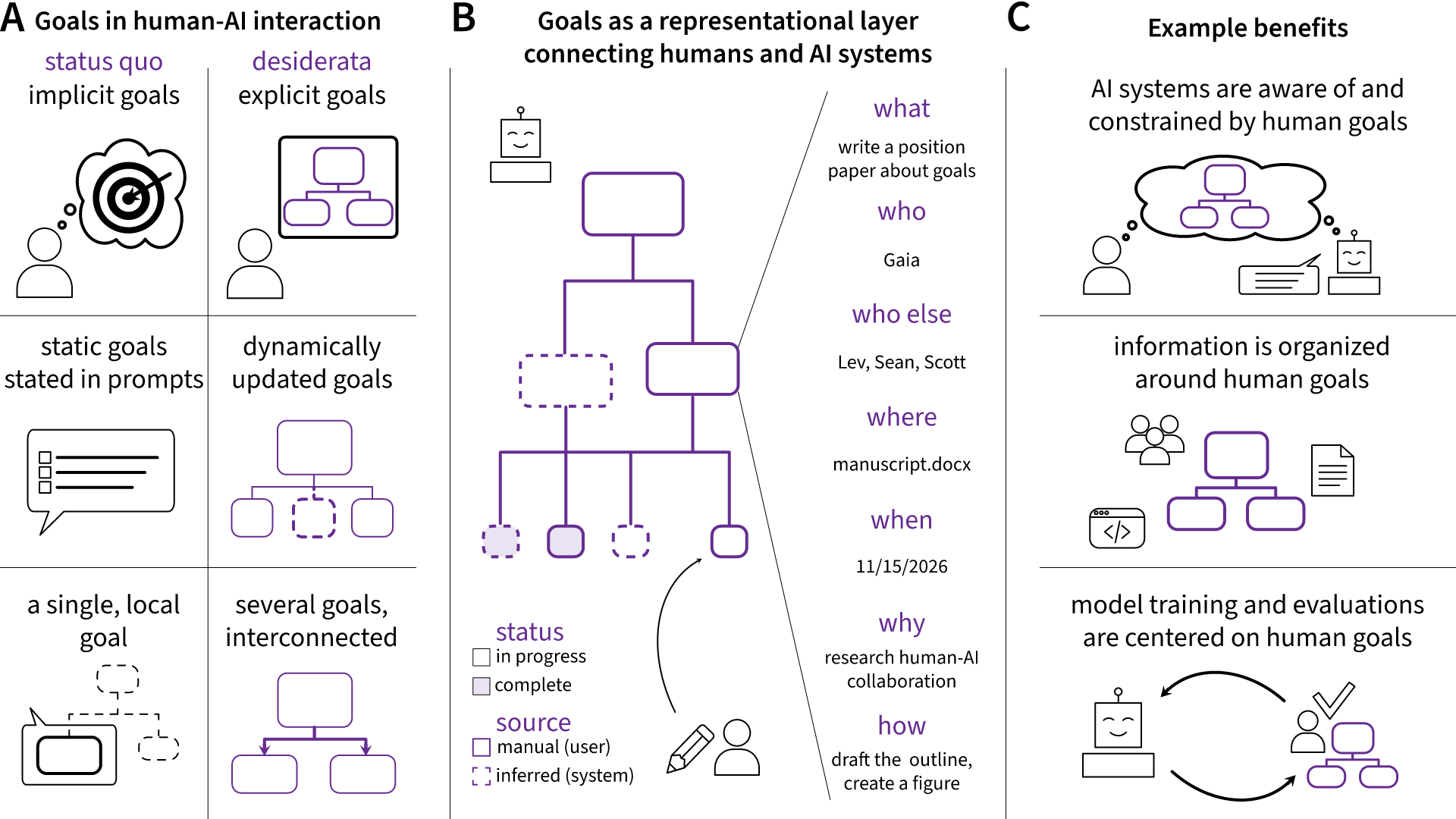}
    \caption{\textbf{Goals as a representational layer for human–AI collaboration.} \textbf{(A)} Implicit, static, or isolated goals motivate explicit, evolving, interconnected representations. \textbf{(B)} An illustrative goal graph links objectives across levels of abstraction, distinguishing endorsed or manually set goals (solid-line) from provisional or inferred goals (dotted-line boxes), and in progress (white-filled) from completed (light purple-filled boxes). An expanded node organizes goal information through seven interrogatives, providing a shared reference for people and AI systems. \textbf{(C)} A representational layer centered around goals could reduce repeated context specification, support information organization, and ground AI evaluation and training in human goals rather than isolated outputs. Other potential benefits, such as improved progress monitoring and organizational coordination, are not shown but discussed in this article.}
    \label{fig:graphical-abstract}
\end{figure*}

Here, we propose goals as an explicit representational layer mediating interactions among people, AI systems, and organizations (Figure \ref{fig:graphical-abstract}). According to this view, goals should be stored as persistent, structured entities that connect requests and activities to the broader outcomes they are intended to serve \citep{tankelevitch2026goals}. Such a layer could link goals to the people, resources, decisions, and artifacts associated with their pursuit, creating a common ground for human--AI interactions organized around the objectives they serve \citep{chen2026conversation}. We focus on human--AI collaboration in knowledge work, while also considering how the same goal representations could support reflection and coordination among human collaborators. Although we do not examine goals outside work, the proposed approach may also be applicable to personal pursuits in domains such as education or health. 

In the remainder of this article, we derive desiderata for a goal layer from established properties of human goals, arguing that effective representations must accommodate their latent, hierarchical, and dynamic nature. We then discuss the opportunities a goal layer could create for context management, reflection, coordination, evaluation, and AI development, before considering practical approaches for constructing goal-centered representations. We conclude by examining the technical and social challenges that must be addressed to realize this vision, including goal inference, alignment, privacy, and governance.

\section{Desiderata for a Goal Layer Follow from the Properties of Human Goals}

A truthful representational layer mediating human--AI collaboration should be built around the properties of human goals themselves, mirroring some of their inherent features while supplementing others. 

\subsection{Goals are partially latent, so goal representations must be explicit}
Many human goals are latent, remaining at least partially unexpressed, unarticulated, or unknown even to the pursuer \citep{bond2008generating}. Individuals may aim toward outcomes they never explicitly communicate, hold abstract goals without a clear specification of what success entails, or discover important objectives only through action and reflection. Goals are often implicit and, even when consciously held, they may not be designed for specificity, challenge, feedback, and commitment \citep{latham1991selfregulation}, or they might fail to clarify when, where, and how goal-directed action will occur \citep{gollwitzer1999implementation}\,---\,necessary features to succeed at actual goal pursuit. This challenge extends to human--AI interaction, where users frequently arrive without fully formed intents and often clarify, revise, or discover their goals through the interaction itself \citep{tack2026llms, yang2026prompts}. Making goals visible and coherent is therefore a necessary first step toward pursuing, coordinating, and evaluating them effectively \citep{slovak2017reflective, tankelevitch2026goals}. 

\subsection{Goals are dynamic, so goal representations must evolve continuously}
Goals are not static targets: they can be refined, reprioritized, and expanded as people acquire new information and take concrete steps towards them \citep{flower1981cognitive, molinaro2026goal}. Evolving social contexts in organizational settings further shape how goals develop throughout their pursuit. Goals may also move through different stages, from setting and active pursuit to abandonment or completion \citep{molinaro2026investigating}. Interactions with generative AI may themselves shape how people formulate and pursue goals \citep{bansal2024challenges, luettgau2025people}. This dynamism creates an adaptive design requirement for goal representation technologies \citep{ekhtiar2023goals}: to support continuous refinement and maintenance, a useful goal layer must be an evolving representation of users' goal ecosystems. 

\subsection{Goals are hierarchical, so goal representations must support abstraction and relationships}
Human goals are inherently hierarchical, ranging from lower-level tasks to broader aspirations \citep{carver1998selfregulation, powers1973behavior}. Psychologically, people can understand the same behavior at multiple levels of abstraction, shifting between concrete actions and higher-level purposes depending on context \citep{vallacher2012action}. Neurally, different brain regions show preferential activation for actions at specific levels of detail \citep{koechlin2003architecture, ribas2011neural}. 
A goal representation layer should thus allow users and AI systems to connect short-term activities to the broader outcomes they support, maintaining all relevant abstractions for context while focusing on immediate targets. 
Moreover, research on goal systems suggests that goals exist within complex networks \citep{kruglanski2018theory}, highlighting relationships such as equifinality, where multiple paths can achieve the same goal, and counterfinality, where actions that advance one objective impede another. Capturing these relationships can provide a richer account of the dependencies and competing priorities that characterize real-world work.

Together, these elements characterize the features needed to accommodate human goals. We next consider what making such representations available to people and AI systems could enable, before turning to practical recommendations and open challenges in their implementation.

\section{Opportunities Enabled by a Goal Representation Layer}\label{sec:opportunities}

Making goal representations available to people and AI systems could create opportunities beyond improving the specification of individual requests. A persistent goal layer could organize context across interactions, support reflection and coordination, and make the purposes of AI-supported work available for scrutiny. It could also provide a reference for studying and evaluating whether AI assistance advances outcomes that people actually value. We expand on each of these possibilities below.

\subsection{Context organization and coordination}
A goal layer could help organize information according to the purposes it serves \citep{yang2026grounding}. Current user prompts often leave implicit why particular requirements matter, if such constraints are specified at all. Once retrieved, goal-relevant information could be used to constrain AI responses. For example, a request to summarize a report may call for different formats depending on whether the user is preparing for a meeting, drafting a proposal, or onboarding to a new project. 
Persistent goal context could also support continuity across work episodes \citep{chen2026conversation}, potentially mitigating AI failures over multi-turn interactions \citep{arike2025evaluating, laban2026llms, park2026authorship}. Moreover, knowledge work involves revisiting prior decisions, responding to changing circumstances, and anticipating future activities \citep{vanukuru2025designing}; beyond providing interaction histories, a goal layer could connect new requests to ongoing activity by identifying which parts remain relevant to the objectives currently being pursued\,---\,thus relieving the user from the burden of restating previously provided information where it is most consequential. 

Shared goal context could also support coordination among participants. Coordination involves managing dependencies among activities
\citep{malone1994interdisciplinary}, including aligning responsibilities, resources, and changing circumstances \citep{schmidt1992taking}.
Goal representations could make the purposes and relationships underlying this work easier to retrieve, compare, and negotiate. For instance, research shows how differing understandings of a meeting's purpose can create tensions around preparation, participation, and expectations \citep{scott2024mental}; making intentions visible could help collaborators recognize where their objectives overlap, diverge, or depend on one another.
This kind of support could extend to collaboration with AI agents. Existing work illustrates how shared, editable plans can support steering and evolving allocations of work within a task \citep{feng2026cocoa}. However, research on communicating long-horizon work finds that people give AI systems more detailed and prescriptive instructions than human collaborators, reflecting expectations of weaker judgment and intent inference \citep{petridis2026compass}. A common layer centered around people's goals could offer a complementary basis for communicating purpose and guidance without requiring as much extra work.
Moreover, centering information around goals could help people discover potential collaborators working towards similar objectives, possibly creating new connections in otherwise siloed organizations. 

\subsection{Reflection and experimentation}
A goal layer could also provide a reference for reflecting on ongoing work.
Indeed, goal pursuit is not always rational \citep{molinaro2026investigating}, but reflective goal setting (wherein workers set current priorities, evaluate their formulation, detail plans to pursue them, and assess progress towards them) promotes productivity \citep{meyer2019enabling}. Experimental evidence also shows that progress monitoring boosts goal attainment \citep{harkin2016monitoring}. By tracking current goals and the steps taken so far to achieve them, a goal layer could support effective monitoring of important milestones, providing opportunities for re-evaluation of previously stated objectives. 
With cognitive offloading to AI posing threats to people's ability to think critically about their own work \citep{lee2025impact}, explicit goal representations could help people question the purposes and assumptions guiding their work, which is particularly relevant as people delegate more execution to AI \citep{sarkar2025agents}. 

Separating desired outcomes from current tasks could also support experimentation at individual, team, and organizational levels. 
Supporting this idea, empirical evidence suggests that starting with the objective can lead to more diverse and effective alternatives for important decisions \citep{siebert2015creating}. Similarly, understanding the core of a problem tends to yield more creative solutions for it \citep{reiter1997problem}. If treated as testable hypotheses about how to achieve an outcome, rather than commitments to a predetermined procedure, goal representations could also enable workers to make more informed decisions about effective courses of action \citep{camuffo2020scientific}. 

\subsection{AI evaluation and development}
Goal representations could also support AI evaluation. By using goals as a unit of analysis to study human--AI collaboration, longitudinal research could examine how assistance affects the objectives people pursue and how they revise them over time. Existing research shows that AI systems shape requirements in ways users do not always recognize \citep{kim2026cotrace}; goal representations preserving authorship could help assess whether people recognize and endorse these contributions. Moreover, while current efforts to estimate the value of AI tools center on output volume, time saved, or the quality of predefined tasks \citep{murphy2026adoption, noy2023experimental}, goals could provide a reference for evaluating AI assistance against outcomes that truly matter to workers, including learning and well-being alongside attributes of key work outcomes \citep{heer2019agency}. Here, discovering more important goals to pursue may itself be a positive outcome despite a lower completion rate. 

Goal-grounded evaluations could, in turn, inform model development. CollabLLM demonstrates that training with rewards estimated over
simulated multi-turn conversations can improve task performance by encouraging models to clarify user intent rather than immediately respond to underspecified requests \citep{wu2025collabllm}. A persistent goal layer could extend this perspective across separate tasks and sessions by linking feedback to the broader objectives that an interaction is intended to advance, thus distinguishing locally satisfactory outputs from contributions that meet a wider set of requirements. Finally, explicit representations of goals and their dynamics could provide additional structure for developing and evaluating human simulators, with current work on modeling users' mental states, including goals, illustrating the viability of this approach \citep{mehri2026goal, wu2026humanlm}. 

\section{Goal-Centered Representations in Practice}\label{sec:practice}

Realizing the opportunities outlined above requires processes for surfacing and maintaining goal information, as well as structures for organizing it. We now consider how these processes and structures could be realized in practice.  

\subsection{Surfacing and updating goals}
Recent work suggests articulation support and inference as complementary mechanisms through which latent goals can be surfaced and dynamically maintained.

Processes for making goals explicit through guided self-reflection have a long history in workplace management \citep{gollwitzer1999implementation, locke2002building, rodgers1991impact} and in modern organizations, technological systems themselves create opportunities for help with goal articulation \citep{bentvelzen2022revisiting, horvitz1999principles}. Conversational AI promises to expand these possibilities \citep{bansal2024challenges, williams2018supporting}. For instance, DiscoverLLM trains AI models to help users turn underspecified preferences into fully formed intents \citep{kim2026discoverllm}. IntentFlow further supports users in expressing their preferences by connecting specific instructions and the outcomes they will yield \citep{kewenig2026metacognitive, kim2026intentflow}. Collaborative planning offers a complementary setting for articulating and refining intentions. Reviewing proposed actions and answering clarifying questions can help users specify requirements before execution \citep{kumar2026plans}\,---\, a process that can be further facilitated by having users and AI agents share an editable representation of tasks, roles, and current progress \citep{feng2026cocoa}. Beyond initial articulation, research on AI-supported writing explores reflective dialogue for revisiting initially declared goals and evaluating whether the emerging text fulfills them \citep{fan2026intention}. 

Goal inference provides an additional source of support. Distributed across text interactions and other user activities, traces of goal-relevant information can provide implicit evidence about what people are trying to accomplish. A longstanding approach treats intent inference as classification into predefined categories, as in identifying informational, navigational, or transactional intentions from search queries \citep{jansen2008determining}, or assigning utterances to supported intent classes in task-oriented dialogue systems \citep{larson2019evaluation}. Generative models expand the ways this task can be approached, making it more suitable for inference at the overarching goal level. Beyond supporting intent classification \citep{arora2024intent}, language models can help discover new intent categories \citep{rodriguez2024intentgpt} and reason about the likelihood of the user's current goals \citep{ma2025flexible}. Multimodal models also enable approaches that generate natural-language descriptions of intended tasks from observed activity, such as general real-world behavior \citep{veerabadran2025benchmarking} and trajectories through web and mobile interfaces \citep{berkovitch2025identifying, gmeiner2025intent}. These capabilities create opportunities for more personalized context-specific goal representations. Inference can also be performed dynamically to update estimates of a person's goal as further actions are observed \citep{zhixuan2024infinite} and detect goal switches \citep{ghose2025changed}.

In an ideal setting, articulation and inference would jointly feed a participatory process of constructing and maintaining goal representations. Although people value an active role in setting their goals at work \citep{couger1990commonalities}, this preference does not necessarily translate into regular reflection on their priorities \citep{baltes2018towards}. Even when people do reflect, the goals they articulate may be shaped by what readily comes to mind due to attentional or emotional biases \citep{shenhav2024affective, tversky1988contingent}, while the alternatives considered may remain limited \citep{beach1990image}. Consistent with this limitation, Bond et al. \citeyearpar{bond2008generating} found that people often omitted objectives they subsequently recognized as important, including in organizational strategic decision-making. Goal inference could thus complement articulation by drawing on distributed evidence to propose candidate goals for people to consider while helping them explore alternative formulations and pathways. 
Inference alone, however, cannot establish a complete representation of what a person wants to pursue, as priorities, constraints, and aspirations may not be visible to the system. The open-ended nature of real-world inference problems only exacerbates this problem \citep{veerabadran2025benchmarking}. On top of improving accuracy, building participation into goal representation systems could also enhance their effectiveness \citep{jorke2023pearl}. In one study, people were more likely to act on goals they set themselves than on AI-generated goals, even when the latter were equally relevant and better formulated, due to stronger psychological ownership with self-formulated goals \citep{chi2026optimized}. Participation also creates opportunities for people to question proposed goals and contest how their representations are used, although realizing these benefits requires privacy and monitoring safeguards, which we discuss in Section~\ref{sec:challenges}. 
These considerations thus favor hybrid approaches in which AI systems help surface and revise candidate goals, while people shape and endorse the actual representations that guide AI assistance.

Current research shows the viability of both articulation support and goal inference mechanisms for goal expression. However, work in both areas largely focuses on the specific task a user is trying to accomplish within an interaction session, offering limited support for understanding how a particular task fits in the broader constellation of projects, priorities, and longer-term aspirations. Bringing these mechanisms together into a goal-centered layer therefore requires connecting immediate intentions to broader objectives and capturing the context needed to support their pursuit. We next introduce one practical framework for structuring this information.

\begin{table*}[t]
\centering
\small
\processtable{Seven interrogatives for structuring goal information,
potential challenges when relevant information is unspecified,
and example benefits of making it available in human--AI interactions.\label{tab:interrogatives}}{%
\begin{tabularx}{\textwidth}{@{}l
    >{\hsize=0.9\hsize\linewidth=\hsize}X
    >{\hsize=1\hsize\linewidth=\hsize}X
    >{\hsize=1.1\hsize\linewidth=\hsize}X
@{}}
    \toprule
    \textbf{Dimension} &
    \textbf{Guiding question} &
    \textbf{Potential challenges if unspecified} &
    \textbf{Example benefits enabled} \\
    \midrule
    What &
    What outcome is being pursued? &
    Actions may lack a clear direction, and progress or completion may be difficult to assess. &
    Ask more targeted clarification questions; establish a shared reference for success; monitor progress. \\

    Who &
    Who is responsible for pursuing this goal? &
    Responsibility may be unclear, making it difficult to determine who should act or be accountable for progress. &
    Adjust assistance to individual workers; remember personal preferences. \\

    Who else &
    Which collaborators, stakeholders, or intended audiences are relevant to its pursuit? &
    Coordination needs or others' requirements may be overlooked. &
    Tailor outputs to intended audiences; account for collaborators' needs and stakeholder requirements. \\

    When &
    When should relevant actions occur? Is there a deadline, duration, recurrence, or planning horizon? &
    The associated goal may lack prioritization. &
    Surface scheduling conflicts; prioritize and organize work; adapt the scope to time constraints; recognize stale context. \\

    Where &
    Which applications, artifacts, tools, or physical and digital spaces are involved? &
    Relevant resources and work contexts may be difficult to connect to the goal. &
    Locate relevant artifacts across applications; ground suggestions in the work environment; preserve continuity when work moves between tools. \\

    Why &
    Which broader objectives does this goal serve? &
    Its contribution to broader objectives may be unclear, making its relevance difficult to evaluate. &
    Interpret requests in light of broader purposes; surface tensions between locally successful actions and wider objectives. \\

    How &
    Through which steps is this goal being pursued? &
    The route from the desired outcome to actionable steps may remain unclear. &
    Translate desired outcomes into actionable subgoals; identify missing steps and dependencies; compare strategies for success. \\
    \bottomrule
\end{tabularx}
}{}
\end{table*}

\subsection{Organizing information in a structured goal graph}
Surfacing goals is not sufficient for an intermediate representation layer to be useful: multiple goals need to be connected to each other, and each should be detailed enough to bridge human intentions and AI work\,---\,comprising information about the people, resources, timing, and broader objectives involved in its pursuit. One simple, yet practical scaffold to structure goal information is provided by seven interrogatives: \emph{what}, \emph{who}, \emph{who else}, \emph{when}, \emph{where}, \emph{why}, and \emph{how}. This structure has precedents in the design and specification of information systems, where similar ontologies have been used to describe enterprise architectures \citep{zachman1987framework} and engineering requirements \citep{vanlamsweerde2001goal}. Here, we propose using these questions as a starting point to organize both descriptive information about individual goals and relationships among them. Making these dimensions explicit also helps identify gaps that may impede or delay goal pursuits: while not all dimensions will be equally important for every goal, absence of information could identify a need for further specification (Table~\ref{tab:interrogatives}).

The first five interrogatives describe a goal and the context of its pursuit. \emph{What} specifies the desired outcome and what success
would look like. \emph{Who} identifies the person or people responsible for pursuing it, while \emph{who else} identifies collaborators, stakeholders, and intended audiences. \emph{When} captures deadlines, durations, recurrence, and planning horizons. \emph{Where} connects the goal to the applications, artifacts, tools, and physical or digital spaces in which relevant work occurs. These dimensions reflect how workplace goals are pursued in practice: establishing shared direction involves not only agreeing on objectives, but also organizing resources, timelines, responsibilities, and collaborative activities \citep{lindley2023building}. Borrowing from action identification theory \citep{vallacher2012action}, the remaining two interrogatives connect goals across levels of abstraction. \emph{Why} links a goal to the broader objective or objectives it serves, making its purpose explicit and spurring motivation \citep{latham1988resolving}. \emph{How} links a higher-level goal to the subgoals by which it can be achieved, enabling effective pursuit \citep{gollwitzer1999implementation}. Together, these relationships form a goal graph. Each linked goal can itself carry descriptive information and connect to further objectives or activities. The graph need not be a tree: multiple pathways may support one goal, and a subordinate goal or activity may contribute to several broader objectives \citep{kruglanski2018theory}. A more advanced representation could also distinguish alternative pathways from components that must be pursued together, perhaps highlighting equivalent goals that can be substituted or discordant goals that await resolution.
The interrogatives can be complemented by metadata that supports ongoing maintenance. A completion status could record whether a goal is being pursued, has been completed, or has been abandoned, while a change history would record revisions to its content and relationships. Provenance information distinguishes user-articulated content from system-generated proposals. Preserving these distinctions also supports visibility into how implicitly captured workplace information contributes to a shared representation \citep{lindley2023building}.

\subsection{Putting goal representations to work}
A structured goal representation does not, by itself, influence how an AI system should assist a person. Putting it to work requires mechanisms for identifying relevant goals, translating their content into guidance, and revisiting that guidance as interactions unfold. While specific implementations will depend on the AI system in question and the particular setting in which it is deployed, some principles should be widely applicable. 

First, systems should selectively connect an immediate request to the goals it may serve. This involves identifying relevant objectives and retrieving enough of their surrounding context to understand the request’s purpose, without treating the entire goal graph as necessary context for every interaction. 

Second, retrieved goal information should inform concrete decisions about assistance. The interrogatives introduced above provide a practical basis for doing so: an intended audience may shape the explanation, a deadline may constrain a proposed plan, and linked artifacts may direct information retrieval. Relationships between goals can also help distinguish desired outcomes from the methods currently being pursued. 

Third, consequential interpretations should remain visible and open to correction. When goal context materially changes the scope of a response or the direction of a plan, the system should make the relevant connection and assumptions inspectable. Clarification is particularly important when plausible interpretations would lead to substantially different commitments or actions. 

Finally, goal-informed assistance should remain responsive to what emerges during the work. New information may change the relevance of an objective, reveal a conflict between goals, or suggest that an existing approach no longer serves its intended purpose. Systems could surface these discrepancies and propose revisions for people to consider. In this way, using the representation also creates opportunities to maintain it: goals inform assistance, while interaction provides evidence through which people can reconsider both their representations and their methods. 

\section{Challenges and Possible Ways Forward}\label{sec:challenges}

While our practical recommendations provide some starting points for building a goal-centered representational layer for human--AI collaboration, they cannot, by themselves, guarantee the success of its deployment. Realizing the proposed opportunities will require overcoming representational, operational, and institutional challenges\,---\,some of which we address below.

\subsection{Inference and validation} 
A first challenge is the inherent difficulty of inferring goals accurately, since neither observed behavior nor natural-language interaction provides direct access to a person's goals. In open-ended settings, this problem is especially difficult because plausible goals need not belong to a predefined set. Moreover, much of the evidence available to a goal inference system is retrospective\,---\,documents created, meetings attended, messages exchanged, and actions taken\,---\,whereas goals are prospective, concerning outcomes a person seeks to bring about. A system that accurately explains what a person has done may therefore still fail to identify what they want to do next. 

These difficulties also complicate evaluation. Ground-truth goal labels are rare \citep{veerabadran2025benchmarking}, and even self-reported goals provide an imperfect reference when people's objectives are incomplete, evolving, or discovered through reflection. Evaluating goal inference therefore requires accounting for uncertainty both in the system's predictions and in the targets against which they are assessed. 
While ongoing work contributes promising approaches for goal inference \citep{ma2025flexible, zhixuan2024infinite}, the participatory approach described in Section~\ref{sec:practice} provides a complementary response to both problems. Rather than treating inferred goals as authoritative representations of users' intentions, systems can surface them as candidates that users revise. These interactions can both improve the representation and provide direct evidence for evaluating inference quality. Over time, such feedback could support evaluation of goal inference in the same evolving context in which the representations are ultimately intended to be used.

\subsection{Maintenance, adaptability, and alignment} 
Even accurately represented goals can quickly become obsolete, requiring ongoing maintenance across timescales ranging from immediate tasks to long-term aspirations. At the same time, asking users to explicitly update or endorse every change could burden them. Systems will therefore need to balance reliance on existing representations, passive updating from new evidence, and requests for user input. 
Goal representations will also need to adapt to the settings in which people operate. Work centered around projects with explicit deliverables and deadlines may support relatively precise representations, while exploratory activities such as research may require goals that remain deliberately provisional. Different roles may likewise produce different goal structures: an individual contributor's work may center on a small number of concrete projects, while managerial work may involve many overlapping objectives distributed across people and timescales. Goal representations should accommodate such variation without imposing excessive structure where ambiguity and flexibility are valuable. 
Moreover, representing a goal is not the same as acting in accordance with it \citep{arike2025evaluating}. Even an accurate and current representation cannot guarantee that an AI system's plans and actions will remain consistent with it. A goal layer should therefore be understood as infrastructure for alignment rather than a solution to alignment itself. 

\subsection{Privacy, authority, and governance} 
Aggregating goal information into a persistent representation could make previously fragmented aspects of a person's work substantially more legible to AI systems and to other people. The same representation that enables personalized assistance could thus also enable excessive monitoring or invasive evaluation. To prevent this, each goal layer must be accompanied by boundaries around its use: not every goal should be visible to every collaborator, manager, organization, or AI system, nor should a goal inferred from behavior automatically acquire the same status as one explicitly endorsed by its owner. A related risk is that goal representations could shift from tools that support people's agency into mechanisms for defining and monitoring what workers should pursue. Automated goal setting is problematic, as AI systems lack human-like task prioritization \citep{molinaro2026language}. As we create tools for surfacing goals, we will have to ensure people have the necessary knowledge and skills to make decisions over them. 

\section{Conclusion}
As AI systems become more deeply integrated into knowledge work, they increasingly operate across workers' activities while lacking a persistent representation of the broader purposes they serve. We have argued for goals as a representational layer that creates common ground in human--AI interactions centered around the outcomes people seek to achieve.
Because human goals are partially latent, dynamic, and hierarchical, useful goal representations must make them explicit while supporting continued refinement, relationships across levels of abstraction, and meaningful human participation in their construction.
Such a layer could provide shared infrastructure for organizing context, coordinating work, supporting reflection and experimentation, and evaluating AI assistance against the outcomes it is intended to advance. Implemented effectively, a goal layer could be a core aspect of the ``AI flywheel'' for knowledge work\,---\,a virtuous cycle where more robust grounding results in better-supported work, which in turn fosters more productive human--AI interactions.  

Realizing this vision will require progress on goal inference and maintenance while anticipating policy and governance issues around its use.
Ultimately, addressing these challenges could provide a foundation for AI systems that are optimized not by the output they produce, but by how effectively they help people pursue and achieve meaningful goals.

\bibliographystyle{plainnat}
\bibliography{ref}

\end{document}